\documentclass{aa}
\usepackage{graphicx}
\usepackage{longtable}
\def\kms{km\thinspace s$^{-1}$}

\def\gs{\mathrel{\raise0.35ex\hbox{$\scriptstyle >$}\kern-0.6em
\lower0.40ex\hbox{{$\scriptstyle \sim$}}}}
\def\ls{\mathrel{\raise0.35ex\hbox{$\scriptstyle <$}\kern-0.6em
\lower0.40ex\hbox{{$\scriptstyle \sim$}}}}

\begin{document}

   \title{The scaling relation of early-type galaxies in clusters. II: 
   Spectroscopic data for galaxies in eight nearby clusters\thanks
{Based on data obtained with the Nordic Optical Telescope (La Palma, Spain) 
with ALFOSC. Also based on observations obtained with DFOSC at the D1.54m 
telescope at the European Southern Observatory (La Silla, Chile)}}

   \author{D. Bettoni\inst{1}
           \and M. Moles\inst{2}
           \and P. Kj\ae rgaard\inst{3}
           \and G. Fasano\inst{1}
           \and J. Varela\inst{1}
           }

   \offprints{D. Bettoni}

   \institute{Osservatorio Astronomico di Padova, Vicolo de'll 
Osservatorio 5, I-35122 Padova\\
              \email{bettoni@pd.astro.it, fasano@pd.astro.it,varela@pd.astro.it}
\and Instituto de Astrof\'{\i}sica de Andaluc\'{\i}a, Consejo 
Superior de Investigaciones Cient\'{\i}ficas, Camino Bajo de Hu\'etor 50, Apdo.  3004, 18080 Granada, Spain\\
              \email{moles@iaa.es}
       \and Copenhagen University Observatory. The Niels Bohr Institute for
              Astronomy, Physics and Geophysics, Juliane Maries Vej 30,
              DK-2100 Copenhagen\\
              \email{per@astro.ku.dk}
                          }

   \date{
         Received
        }
\titlerunning{Spectroscopic data for galaxies in eight nearby clusters}
\abstract{}
{We present in this work low and intermediate resolution spectroscopic data
collected for 152 early type galaxies in 8 nearby clusters with z $\leq$ 0.10.} 
{We use low resolution data to produce the redshift and the K-correction 
for every individual galaxy, as well as to give their overall spectral energy 
distribution, and some spectral indicators, including the 4000\AA~break, the
Mg$_2$ strength, and the NaD equivalent width.We have also obtained higher 
resolution data for early type galaxies in three of the clusters, to determine 
their central velocity dispersion. }
{The effect of the resolution on the measured parameters is discussed.} 
{New accurate systemic redshift and velocity dispersion is presented for four of 
the surveyed clusters, A98, A3125, A3330, and DC2103-39. We have found that the
K-correction values for E/S0 bright galaxies in a given nearby clusters are very 
similar. We also find that the distribution of the line indicators significantly 
differ from cluster to cluster. } 
          
\maketitle
\section{Introduction}

This is the second paper to present data on nearby clusters of galaxies. 
The motivations and details of the program have been given in Fasano et al 
(2002, hereafter Paper I; see also Fasano et al 2000), following the proposal 
discussed by Kj\ae rgaard, J\o rgensen, and Moles (1993, hereafter KJM; see also 
Moles et al. 1998). They will be used together with the results from the 
project WINGS (Fasano et al. 2006) to analyze, among others the scaling relations
of early-type galaxies in clusters.

Here we present low resolution (20\AA~and 40\AA, 
LRS/20 and LRS/40 respectively in the following) and intermediate resolution
spectroscopy (3.3~\AA~resolution, IRS in the following) of 152 early type 
galaxies in 8 nearby clusters. The LRS data are intended to provide the redshifts 
of the candidate galaxies, to establish their membership to a given cluster. 
They also allow the measurement of the K-correction for each individual galaxy 
and the strength of some spectral features. Apart from the more often used Mg$_2$,
we consider here two other prominent spectral features, the NaD doublet and the 4000\AA~break, D4000. The NaD line 
index is more sensitive to temperature effects than Mg$_2$, but it can be affected
by the presence of interstellar material (Burstein et al 1984; Bica, and Alloin 
1986; Bica et al 1991). However, once allowance is made for that, it correlates 
well with other indices, Mg$_2$ in particular. Bica, and Alloin (1986) concluded 
that for a metallicity not greater than solar, any excess of NaD (with respect to
the prediction  from the Mg$_2$ indicator) can be due only to interstellar 
absorption. An outlying position in the Mg$_2$-NaD plane can therefore be 
interpreted as a sign of peculiarity (see Bica et al 1991).


 \begin{table}
      \caption[]{Log and setup of the observations}
         \label{log}
      $${\footnotesize
         \begin{array}{ccccrcccc}
            \hline\noalign{\smallskip}

{\rm Run} & {\rm Date} & {\rm Mode} & {\rm Tel.} & {\rm \arcsec/pix}
& {\rm Gr.} & {\rm \lambda} & {\rm \AA/pix}  &  {\rm Res } \\
& & & & & \# & range(\AA)& &\AA^{\mathrm{a}}\\
\hline\noalign{\smallskip}
1 & {\rm  Dec/94}& LR & D1.54 &  0.49 &  4 & 3300-6400 &  3.9 &  20 \\
2 & {\rm Sep/95} & LR & D1.54 &  0.49 &  4 & 3300-6400 &  3.9 &  20 \\
  &              & IR & D1.54 &  0.49 & 13 & 4800-5800 & 0.95 &  3.3 \\
3 & {\rm Feb/97} & LR & NOT   & 0.188 &  7 & 3800-6800 &  1.5 &  20 \\
4 & {\rm Aug/98} & LR & NOT   & 0.188 &  4 & 3300-6400 &  3.1 &  41 \\
5 & {\rm Aug/99} & LR & NOT   & 0.188 &  4 & 3300-6400 &  3.1 &  41 \\
6 & {\rm Mar/01} & IR & NOT & 0.188 & 13 & 4800-5800 &  0.5 &  1.4  \\

            \noalign{\smallskip}\hline\noalign{\medskip}
 \end{array}}
     $$
\begin{list}{}{}
\item[$^{\mathrm{a}}$] Resolution corresponds to the 2\farcs5 slit used in all the
runs
\end{list}
   \end{table}
   

The 4000\AA~break, D4000, is primarily sensitive to the presence of young stars. 
Hamilton (1985) selected it for that reason as a mark to study the evolution of 
early type galaxies with redshift. Dressler and Shectman (1987) concluded that 
it is not sensitive to metallicity but only to the presence of young stars, and 
insisted on the adequacy of the break indicator to follow the cosmic evolution of 
early type galaxies. On their side, Kimble et al (1989) found that D4000 does 
correlate with some metallic indicators. The theoretical work by Poggianti and 
Barbaro (1997) showed that the break is strongly dependent on the effective 
temperature, and is also sensitive to the metallicity, but only for intermediate 
temperature stars. On this basis, Barbaro and Poggianti (1997) have developed 
evolutionary models showing that D4000 would be a  measure of the present star 
formation rate. The calibration of D4000 in terms of the atmospheric stellar 
parameters is rather complicated as many absorption lines are included in the 
break. Gorgas et al (1999 and references therein) have made an  
empirical calibration of the break that can be incorporated into the evolutionary
models to predict its value. The most interesting aspect here of the break 
indicator is its sensitivity to the recent star formation, and therefore its 
ability to trace evolution.

The intermediate resolution spectroscopy (IRS) data are necessary to determine 
the central velocity dispersion needed to build the Fundamental Plane. They also 
provide more accurate spectral line indices, within the covered spectral range. 
Unfortunately the behavior of the CCDs we used was very noisy at wavelengths 
shorter than ~5200\AA. Thus the measurement of the H$\beta$ and other indices 
related to rather weak features was uncertain and therefore we do not include 
them in the present work.

\section{Observations and Data Reduction}
The observations were carried out during several runs since 1994 with the 
Danish 1.54m telescope (DKT) at La Silla (Chile), and the Nordic Telescope (NOT)
at La Palma (Spain). We used identical focal reducer instruments with both 
telescopes, DFOSC (DKT) and ALFOSC (NOT), equipped with identical grisms. 
Some of the data were taken using the MOS possibilities with ALFOSC. 
The different runs and instrumental setups are given in Table~\ref{log}. 
In all observations we used a 2\farcs5 slit. For the first run the detector 
was a Thompson 1024$\times$1024 CCD with 19$\mu$m pixels, whereas we used various
thinned Ford-Loral 2024$\times$2024 CCD's with 15$\mu$m pixels for the other runs.

Template stars, for the measure of radial velocities and the velocity 
dispersion, of spectral type G8-K3 III, were observed each night as well as 
nearby galaxies with known and accurate data. These last were used as standard 
to gauge the accuracy of the calibrations. We also observed several flux standard 
stars to allow the data to be fully calibrated.
Identification and positional information for the cluster galaxies are taken 
from Dressler (1980). Otherwise we measured the positions on the DSS. In 
Table~\ref{pos} in column 1 we give the name (from NED or Simbad databases) 
in column 2 and 3 the coordinates and finally in column 4 we give the reference
identification we used for the spectrum, this identification is that used in 
Tables~\ref{res} and \ref{finalres} to identify the galaxies for clusters A98 and A3330.
The identifications in Tables~\ref{res} and \ref{finalres} for galaxies in the 
remaining clusters are from Dressler (1980).

The data reduction procedure was similar for the LRS and IRS data. It was 
performed within IRAF\footnote{IRAF is the Image Analysis and Reduction Facility 
made available to the astronomical community by the National Optical Astronomy 
Observatories, which are operated by the Association of Universities for Research
in Astronomy (AURA), Inc., under contract with the U.S. National Science 
Foundation.} the scientific frames were corrected for bias and flat-field and
calibrated using He/Ar arc-lamp. 
For the MOS data we extracted individual 2D spectra for each object and 
treated as any other spectra, so we don't refer to them explicitly in the 
following.

\subsection{The Measurement of the Redshift. Internal and external accuracy}

Before attempting to measure the redshift of the target galaxies, we checked 
the accuracy of the zero point in the $\lambda$-calibration of every spectrum,
examining the position of the night sky line [OI]$\lambda$5577.32 \AA. When 
necessary the spectrum was shifted to that nominal position. For the radial 
velocity standard stars, given the short exposure time, the night sky lines 
were not detected so this procedure could not be applied. We decided to put the 
standards at zero velocity using their own spectral lines; then we used the IRAF
cross-correlation package to control the consistency of the method, i. e., that
together, they all define a zero velocity system, within the uncertainties. 

The standard stars were then used as templates to measure the redshift of the
standard galaxies, using the same IRAF packages. The radial velocity was 
determined as the mean of the results for the different template stars. The 
typical scatter amounts to 15~\kms. In all cases the scatter was within the 
formal uncertainties of the parameters. The results for those standard galaxies
are given in Table~\ref{res}. 

Comparing the results obtained for the standard galaxies from LRS/20 and IRS 
data (9 cases; there are not galaxies observed in both IRS and LRS/40 modes) we 
find that the differences in redshift have an average of 7~\kms,  with a scatter
of 32~\kms. Part of that scatter is due to just one galaxy, E462G15, for which 
the IRS result is 71~\kms~higher than the LRS one. Notice that the IRS value is
much closer to the redshift reported by J\o rgensen, Franx \& Kj\ae rgaard (1996, 
JFK) for that galaxy. 

To check the external consistency of the data we have compared them with other 
sources, JFK (10 galaxies in common, 7 with IRS data) and Smith et al (2000; 7 
objects in common, 5 with IRS data). If we compare only our IRS data, the 
agreement is excellent with both sources, with $\sigma$(diff.)$\sim$ 15~\kms. 
Indeed, the comparison is also good when LRS data are included, except for
EG462G15. 


Once the consistency of  our measured redshift was assured, we decided 
to use the standard galaxies as templates to determine the redshift of the target 
cluster galaxies. We found that they give more accurate results than the standard 
stars, due to the better spectral matching between both sets of galaxies. Each 
target object was cross-correlated with all the templates. The average of the 
resulting z values was taken as the redshift of the galaxy, and the scatter as a
quality indicator. The heliocentric redshift values are given in Table~\ref{res}. 
The code 1 to 4 corresponds to 1$\sigma$ values of $\leq$50~\kms, between 50 and
100~\kms, between 100 and 200~\kms, and between 200 and 300~\kms, respectively.
That scatter is typically smaller than 75~\kms, with more than 75\% of them within 
150~\kms, even if it reaches up to 300~\kms~ in some particular cases.

\begin{figure}
\centering
\includegraphics[angle=-90, width=8.5cm]{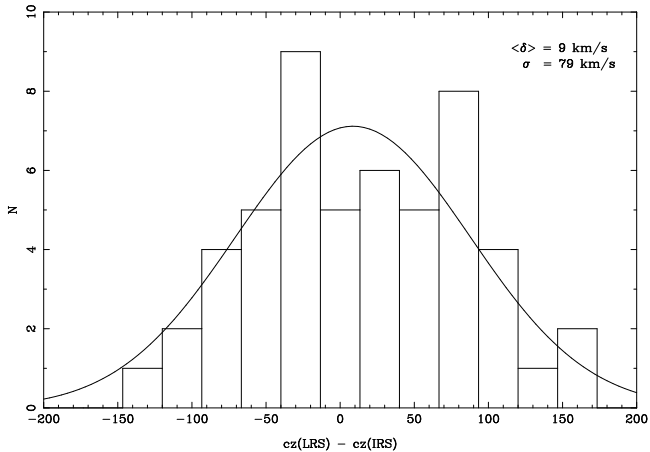}
\caption{Distribution of the cz differences for the 52 objects that
have both IRS and LRS data. The bin size is 25~\kms. The solid line is 
the gaussian fit to the binned data.}
\label{compaz}
\end{figure}

The internal accuracy of the LRS data can be assessed looking at
Figures~\ref{compaz} and \ref{compaMg}.

%
%

We have searched the literature to examine the external accuracy of our LRS data.
The comparison was restricted to those sources for which there is a minimum of 5 
galaxies in common. The agreement is in general good as can be seen in 
Table~\ref{compacz}. Looking to the actual figures in the
table, it would seem that our cz values tend to be slightly greater than most of 
the  others. Whereas this cannot be excluded, the differences are about 50~\kms or
smaller, so we can conclude that our redshift measurements are in the same system
as those reported in the quoted references. Indeed, the situation is similar for 
the IRS data, (see Table~\ref{compacz}). 
The cases for which the disagreement is higher and cannot be 
accounted for by mere observational errors have been flagged in 
Table~\ref{res}, and commented in the footnotes.

\subsection{The Measurement of the central velocity dispersion}

The velocity dispersion of each target galaxy was determined as the mean of the 
results for the different template stars. The typical scatter is 10~\kms. In all 
cases the scatter was within the formal uncertainties of the parameters. The
velocity dispersion and the radial velocity were derived, both for the comparison
galaxies and for the target galaxies, with the Fourier Quotient Technique 
(Sargent et al 1977; Bertola et al 1984), using the standard stars as templates.
The results are presented in Table~\ref{res}. Comparison with the JFK values 
shows a good agreement. The differences have an rms value of 15~\kms. 
Unfortunately the external checking can only be done for a handful of target 
galaxies. We have found independent determinations for only 5 of the galaxies 
in our sample, all from Wegner et al (1999). For them, the agreement is also 
satisfactory since the rms of the differences is the same as for the standard 
galaxies, 15~\kms.

%
\begin{table}
      \caption[]{External comparison of the redshift data}
         \label{compacz}
$$
         \begin{array}{l|cccc|ccc}
            \hline\noalign{\smallskip}

{\rm Cluster} & \multicolumn{4}{c}{\rm LRS} &  \multicolumn{3}{c}{\rm IRS} \\

              & {\rm Ref.} & {\rm N} & {\rm \Delta(cz)} & {\rm \sigma}
                           & {\rm N} & {\rm \Delta (cz)} & {\rm \sigma}\\
\hline\noalign{\smallskip}
A98   & 1 & 10 & $+$36 & 147 &  -- & --  & -- \\
      & 2 & 11 & $-$10 & 120 &  -- & --  & -- \\
A119  & 3 & 14 & $+$52 &  75 &  11 & 80  & 48 \\
      & 4 &  6 & $+$32 &  91 & 6   & 46  & 57 \\
      & 5 &  5 & $+$3  &  56 & 5   & 28  & 52 \\
A3125 & 8 &  6 & $-$11 & 121 & 5   & $-$17 & 133 \\
A1069 & 3 &  7 & $+$29 & 116 &  -- & --  & -- \\
      & 6 &  6 & $+$15 & 162 &  -- & --  & -- \\
A1983 & 7 & 14 & $+$31 &  77 &  -- & --  & -- \\
A2151 & 7 &  9 & $-$15 & 153 &  -- & --  & -- \\

\noalign{\smallskip}\hline\noalign{\medskip}
\end{array}
     $$
\begin{scriptsize}
References. 1, Beers et al (1982); 2, Zabludoff et al (1990); 3, Katgert et al (1998); 4, Wegner et al (1999); 5, Huchra et al (1999); 6, Beers et al (1991); 7, Dressler \& Shectman (1988); 8, Caldwell, and Rose (1997).
\end{scriptsize}
   \end{table}

\subsection{The Measurement of the K-effect and of the spectral features}

To measure the spectral characteristics, the spectra were first shifted to zero
redshift, using the measured cz values as we mentioned, with the appropriate 
tasks within IRAF. Then, the K-effect was measured by computing the magnitude in 
a given band in both the observed and the zero-redshifted spectra. Given the 
spectral coverage of our data and the redshift of the sources, it was not always 
possible to compute the K-effect for all the three B, V and Gunn~r bands. 
In fact, the only band for which we have measured the K-effect for all the
targets is V. The results are given in Table~\ref{res}.

In Table~\ref{rescum} we present the average values for every cluster we have 
observed. Comparing with the values reported by Pence (1976) for E/S0 galaxies, 
and with the model predictions by Poggianti (1997) for the same type of galaxies,
we find a very good agreement. We notice that the scatter of the K-corrections 
for early type galaxies in clusters is very small, indicating that applying the 
same correction to all the bright E/S0 galaxies in clusters introduces only small 
errors.

To measure the 4000~\AA~break we choose to use bands of 100~\AA~ width on both
sides of the feature. This choice is just practical, due to the fact that our 
spectra do not extend enough into the blue. The data are also collected in 
Table~\ref{res}. Even if we cannot directly compare with similar measurements 
from other authors, we have verified that the range of our values is totally 
similar to that found by Dressler \& Shectman (1988) for similar galaxies in 
nearby clusters. For the different line indicators we have used the Lick 
spectral bands as defined in Worthy et al (1994). For reasons we have already 
indicated, only the most prominent features, Mg$_2$, NaD, and the 4000\AA~ break 
were measured. The results are presented in Table~\ref{res}. 

Let's start considering the Mg$_2$ strength measurements. First, the internal
accuracy can be ascertained from repeated measurements with the same resolution. 
The rms of the differences for galaxies observed twice in the LRS/20 mode amounts
to 0.014, most probably due to the poor S/N of some of the spectra we repeated. 
Thus, as for the redshift, we consider that the overall quality of the data is 
definitely better than that figure. Then for the standard galaxies we find a good
agreement with JFK values, even if the resolution used is not the same. 
We notice that the same aperture was used by JFK and here.  Except for N1403, 
for which the Mg$_2$ value was quoted as uncertain by JFK, as is our own value, 
due too rather poor S/N, the rms value of the differences is slightly smaller 
than 0.01 mag, i. e., smaller than for repeated measurements.  As we are going 
to discuss below, this  difference mainly reflects the effect of the resolution. 
Incidentally we notice that the comparison shows up that our results are in the
same system as those reported by JFK.

\begin{figure}
\centering
\includegraphics[angle=-90, width=9cm]{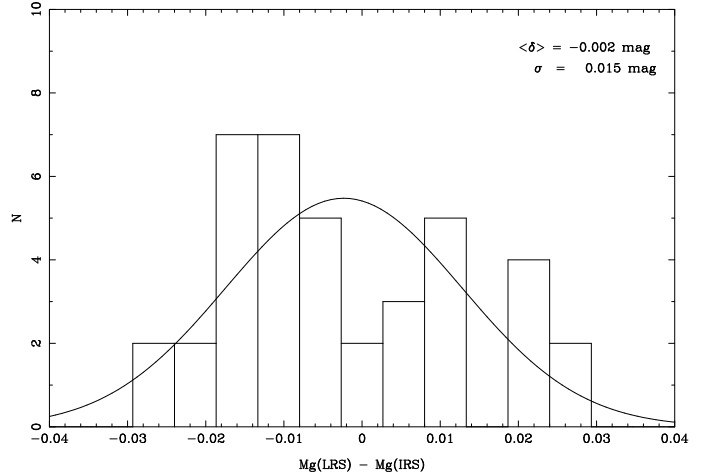}
\caption{Distribution of the differences in the Mg$_2$ values for 42 objects
for which both IRS and LRS data are available. The bin size is 0.004. 
The solid line is the gaussian fit to the binned data}
\label{compaMg}
\end{figure}

The effect of the resolution on the measured values can be checked comparing 
the results from LRS/40, LRS/20 and IRS data. Starting with the standard 
galaxies, for the 8 objects measured in LRS/20 and IRS modes, the average 
difference amounts to 0.004 mag, with an rms value of 0.014 mag. Much of 
that scatter is due to a single object, N1395, for which the LRS value is 
0.028 mag smaller than the IRS value. Regarding the target galaxies, 
we have 42 objects for which we could measure the Mg$2$ strength from LRS and 
IRS data. Excluding three grossly discrepant cases (more than 0.03 mag 
difference), the differences have an average of 0.003 mag, with $\sigma$ =  
0.014 mag. Their median value amounts to 0.006 mag (see Figure~\ref{compaMg}). 
%

\begin{figure}
\centering
\includegraphics[angle=-90, width=9cm]{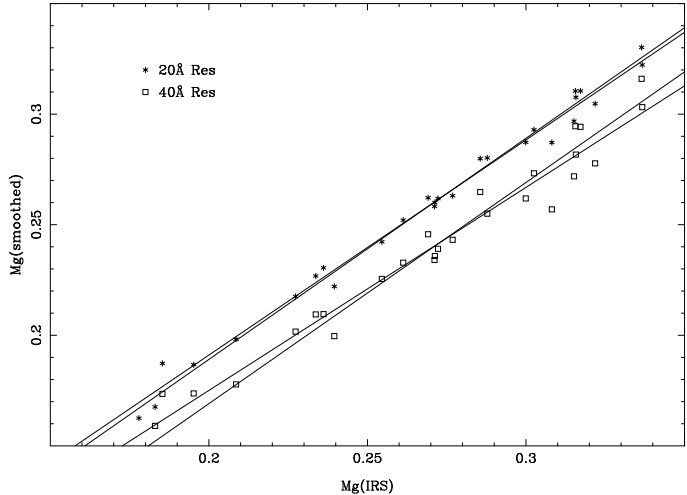}
\caption{The  relation between the Mg$_2$ strength values obtained from 
different resolution data. The plotted lines for the two sets of data are the 
best fit and the fit with slope = 1 in each case.}
\label{Mgresol}
\end{figure}

To test further the effect we have degraded the observed IRS data to produce 
spectra with the resolution of the LRS/20 and LRS/40 modes. The results are 
presented in Figure~\ref{Mgresol}. It can be seen that the relations have slopes
very close to 1, in particular  between LRS/20 and IRS  data. The systematic 
effect can be represented by the relation Mg$_2$(IRS) = Mg$_2$(LRS/20) + 0.011
mag, with a scatter of 0.004 mag. For the LRS/40 data the correction amounts 
to 0.031 mag, with a scatter of 0.007 mag.
%

\begin{figure}
\includegraphics[angle=-90, width=9cm]{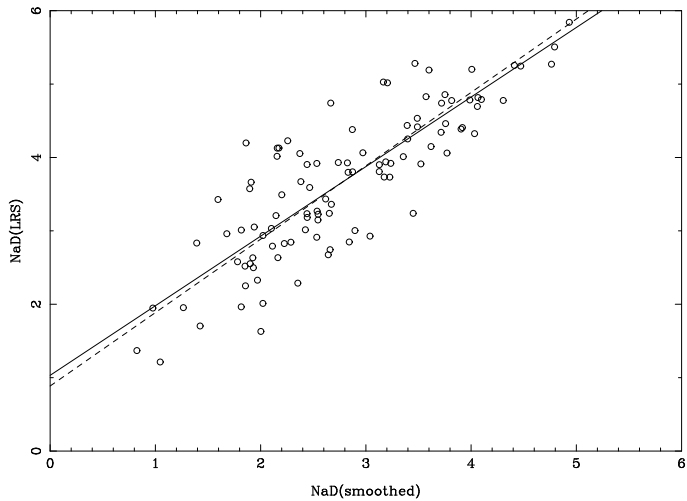}
\caption{The EW of the NaD line from real LRS/20 data and from the same spectra 
degraded to 40~\AA~ resolution. The lines are the best fit and the fit with slope 
= 1. }
\label{NaDresol}
\end{figure}

For the other line indicators, we estimate that the errors are of the order of
5\% at worst, a figure that is confirmed by the comparison of the values obtained
from repeated measurements. The 4000~\AA~break is a very robust indicator that is 
not affected by other aspects, whereas an important resolution effect is expected
on the NaD values since the Lick bands used to define the index are rather narrow. Given that the wavelength range of the IRS data does not cover the NaD region, and we have not enough data taken in both LRS modes for a sound comparison, we have simply measured the NaD equivalent width for the LRS/20 data degraded to the resolution of the runs 4 and 5, namely 40~\AA. The results are plotted in Figure~\ref{NaDresol}. We find that the correction is of the form EW(LRS/20) = EW(LRS/40) + 0.9 ~\AA. This is the correction we will apply to put all the NaD EW values in the same LRS/20 system.

%

\subsection{The aperture corrections: The final data}

To perform the aperture correction of the Mg$_2$ index we have followed JFK to
transform the observed values to the standard aperture of 1.19h\thinspace$^{-1}$ 
kpc, corresponding to 3\farcs4 at the distance of Coma. Their expression (4) was
used to evaluate the correction. As we have already noticed, our data are in a 
system similar to that of JFK, so we have still to transform our values to put 
them on the Lick system. This has been done also after JFK, adding 0.011 to our 
values. Finally, we have not corrected for the velocity dispersion simply because 
it is not known for most of our galaxies; in any case, for a correction as that 
proposed by JFK, it would never be bigger than 0.003.

Similarly, an aperture corrections should be applied to the other line indicators.
For the NaD equivalent width, given that there are no specific data on its radial 
gradient, we decided to apply the same kind of correction as for the Mg$_2$ index.
The final expression for the aperture corrected NaD line EW is given by
$$ (EW)_n = W[1- (r_o/r_n)^{-0.016}] + (EW)_o\times (r_o/r_n)^{-0.016}$$
\noindent 
where the index n(o) stands for aperture corrected (observed) value, and W is the
width of the filter used to measure the feature (32.5 \AA~ in the system defined 
by Worthy et al 1994). The maximum correction, for A98, amounts to 0.60\AA, 
whereas it is of 0.18\AA~ for A119.

For the aperture correction of the 4000\AA~break we have used the result by 
S\'anchez-Bl\'azquez et al (2001), who have found that the break changes as 
$-$0.20log(r). The maximum correction amounts to 0.11.

The final, corrected values for the redshift, the velocity dispersion and the 
line indicators are given in Table~\ref{finalres}. We have taken the Mg$_2$ 
results from the IRS mode when available, correcting the values obtained with 
LRS data to that resolution following the recipes explicited before, including 
the correction to put them in the Lick system. For the NaD EW value, we have 
corrected all the LRS/40 measurements to LRS/20 just adding 0.90\AA~ to take 
into account the resolution effect as discussed before.

\section{General Considerations}

\subsection{The Redshift of the Clusters}

For four clusters, namely A119, A1069, A1983 and A2151, the number of new 
(or significantly modified) redshift values we can add represents only a small 
fraction of the total known. There are some cases however, namely A98, A3125,
A3330, and DC2103 where that number is significantly increased. Therefore, a new
and more precise determination of their redshift and velocity dispersion is 
possible. We discuss them briefly below. The results are given in 
Table~\ref{rescum}.

\begin{figure}
\includegraphics[ width=9cm]{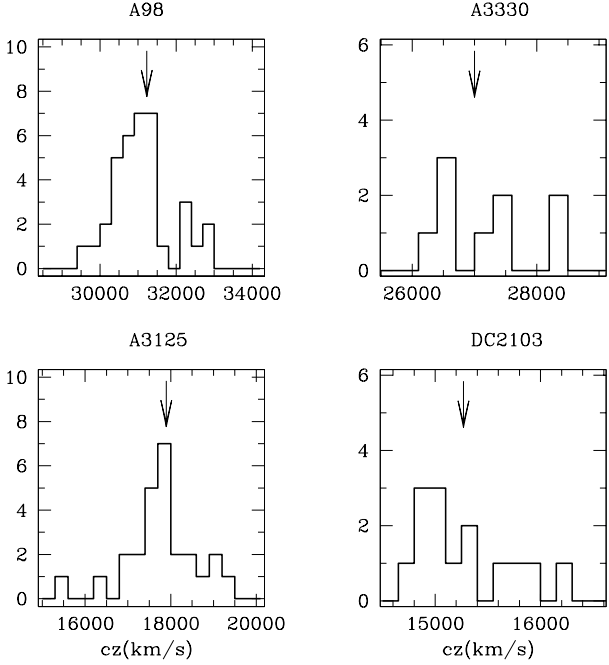}
\caption{Distribution of known redshift of galaxies in the clusters A3125, 
DC2103, A3330 and A98. The arrows indicate the cluster mean cz value as given 
in the text. }
\label{redshift}
\end{figure}

We have obtained the redshift for 31 galaxies in the field of A98 
(see Table~\ref{rescum}), of which 16 are also in Beers et al (1982), 
or in Zabludoff et al (1990), or in both. Following the preceding discussion 
on the redshift accuracy of our data, we have adopted our values in all cases
when they were also in other references. The redshift distribution of all the
galaxies with known cz in the field of A98 is presented in the 
Figure~\ref{redshift}. Considering all the 38 objects with cz in the range 
between 29500~\kms~ and 33000~\kms~ , we find for the cluster redshift cz  
= 31225~\kms, with $\sigma$ = 895~\kms. We notice that 6 of the 7 foreground 
galaxies define a group with cz = 17973~\kms, and $\sigma$ = 254~\kms.
%

%

Beers et al (1982) argued that A98 would comprise two spatially and dynamically 
distinct extensions. With the new data, a total of 11 galaxies are located in the 
A98N extension, and 25 in A98S. The velocity distribution shown in 
Figure~\ref{redshift} shows only one peak and is rather smooth. Looking at the
two extensions,  there is a slight redshift difference between the northern and 
southern extensions, but is not significant. Therefore, the existence of two 
substructures in cz space is not confirmed with the still scarce data already 
available. 

Considering A3125, Caldwell and Rose (1997) have reported the redshift of 16 
galaxies in the field of the cluster. Only five are also in our sample. Together 
with our measurements, the total amounts now to 31  values. The cz distribution
is given in Figure~\ref{redshift}. Taking out the 5 outliers (2 fore- and 3 
background galaxies), we are left with 26 objects, which define the cluster 
redshift cz = 17898~\kms, with $\sigma$ = 779~\kms.

The redshift quoted for A3330 in the compilation by Struble and Rood (1999), 
z = 0.0921, was determined with only two galaxies (Ebeling and Maddox 1997). 
From the 9 galaxies we have measured, we  find cz = 27000~\kms, and $\sigma$ = 
695\kms, with the distribution plotted in Figure~\ref{redshift}. 

The situation is similar for the cluster DC2103-39, for which only 3 redshifts
were known (Loveday et al 1996). We have measured the redshift of 21 galaxies 
in the area, of which 6 are foreground and 1 is a background object . For the 
remaining 14 objects (see Figure~\ref{redshift}) we find cz = 15268~\kms, 
with $\sigma$ = 449~\kms. All the six foreground objects are at a similar 
redshift. Five of them are grouped with an average cz = 9260~\kms,
with $\sigma$ = 70~\kms. These galaxies are located in the field we called 
``a'' in Paper I. They could be part of a cluster as more galaxies are visible 
in that field. 

\subsection{The range of spectral properties}

In Table~\ref{rescum} we also give the range and median values of the 
K-correction terms and the spectral indicators we have considered here. The 
K-correction values, as indicated before, are very similar for all the bright 
E and S0 galaxies in a given cluster. The average values are very well defined, 
with a small scatter, and follow the trend with z shown by the data presented by
Pence (1976) and with the model prediction by Poggianti (1997) for the same 
kind of galaxies.

\begin{figure}
\includegraphics[angle=0, width=9cm]{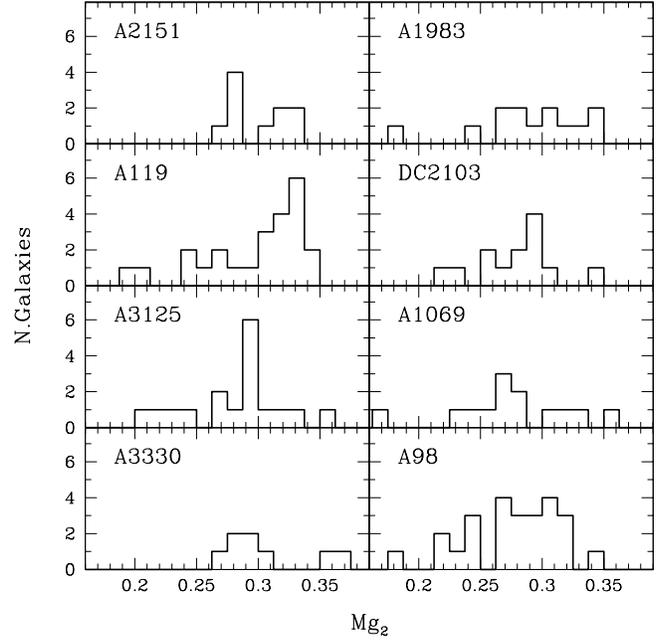}
\caption{Distribution of the  Mg$_2$ strength for all the measured galaxies in 
each cluster.}
\label{MgCum}
\end{figure}
%
\begin{figure}
\includegraphics[angle=0, width=9cm]{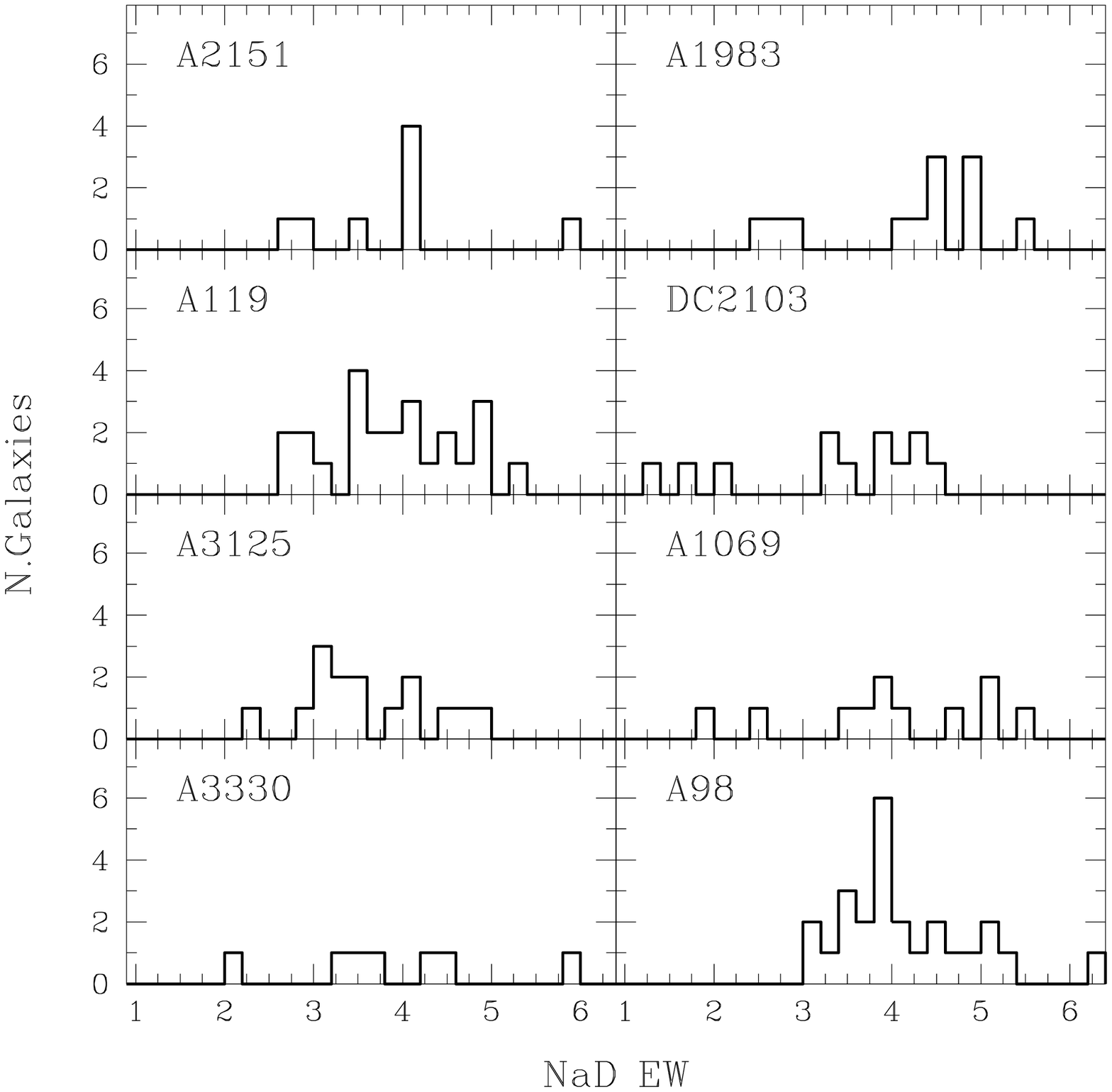}
\caption{Distribution of the NaD equivalent width for all the measured 
galaxies in each
cluster. }
\label{NaDCum}
\end{figure}
%
\begin{figure}
\includegraphics[angle=0, width=9cm]{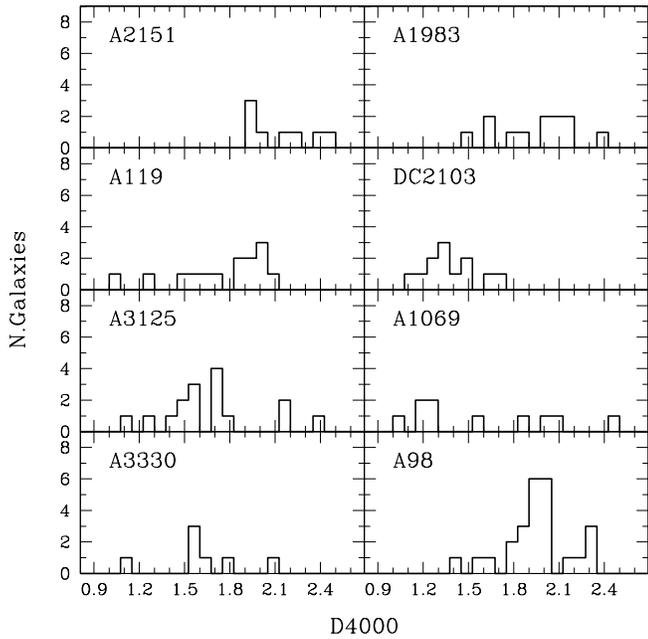}
\caption{Distribution of the the 4000\AA~break line indicator for all the 
measured galaxies in each cluster.  } 
\label{D4000Cum}
\end{figure}

\begin{figure}
\includegraphics[angle=-90, width=8.5cm]{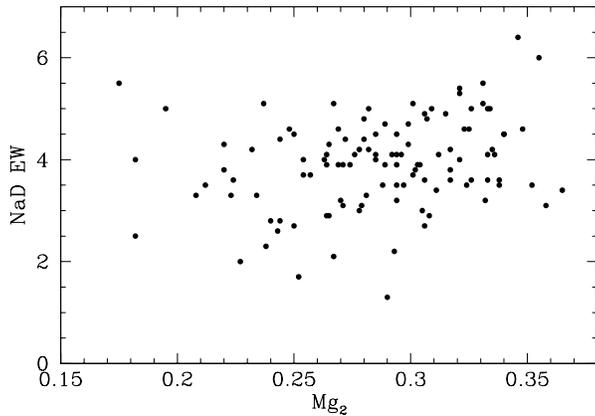}
\caption{The Mg$_2$ strength {\sl versus} the NaD equivalent length for the
galaxies in different clusters}
\label{MgNaD}
\end{figure}

The Mg$_2$ values span a rather big range in all the 8 clusters 
(see Figure~\ref{MgCum}), in most cases from ~0.20 to ~0.35. The exceptions 
are A2151 and A3330 with higher minimum values, what could be due to the small
number of objects observed in these clusters.  
The median values are however different, from 0.269 mag in A1069 to 0.320 mag 
in A2151. There is no significant trend with cz, finding clusters at similar cz 
with different median values. The situation is similar for the NaD 
(Figure~\ref{NaDCum}) equivalent width, with also spans a big range of values 
in each cluster.  The median values are however within 1\AA. Finally, for the 
4000~\AA~ break (Figure~\ref{D4000Cum}), there are differences in the ranges, but 
could be simply due to the small number of sources. In any case we point out the
absence of low D4000 values in A2151, and that of high values in DC2103. 
We notice that a trend is visible between the NaD and Mg$_2$ indicator, 
but with an important scatter (see Figure~\ref{MgNaD}).

\subsection{Galaxies with emission lines}

The targets we selected were among the brightest cluster galaxies classified 
as E or S0, so it was not expected to find but a few showing emission lines. 
Out of a total of 13 of such objects objects, half are actually foreground 
objects relative to the cluster under consideration. These are A1069d41, 
A3125d11, A3125d14, A98F3g4, and DC2103D18. They all have typical HII-like 
spectra.

The cluster member galaxies with emission lines are the following:

\begin{itemize}
\item{A119d44. Faint [OIII]$\lambda$5007 detected only in the IRS data}
\item{A119d45. This galaxy presents a spectrum typical of a star forming region}
\item{A3125d160.Faint [OIII]$\lambda$5007 detected only in the IRS data}
\item{A3125d77. It presents strong H$\alpha$ emission, with some structure. 
There is some indication of high electronic density from the [SII] lines. 
Neither the [OIII] lines, nor H$\beta$ are visible in our spectrum}
\item{A2151d78. This galaxy (IC1182) is a known peculiar early type object, 
with a spectrum in the border line between star forming and active galaxies. 
Its properties have been discussed in Moles et al (2004)}
\item{A98F1g44. It presents all the lines typical of star forming galaxies, 
from [OII] to [SII]}
\item{A98F1g58. Spectrum typical of an early type galaxy, but with H$\alpha$,
[NII] and [SII] clearly visible}
\item{DC2103d18. Faint [OIII]$\lambda$5007 detected only in the IRS data}
\end{itemize}

We notice that all have spectra of star forming regions, and none shows signs 
of nuclear activity.

\section{Summary and Conclusions}

We present here LRS and/or IRS for 147 early type galaxies in 8
nearby clusters with z $\leq$ 0.1. Data on the redshift, velocity dispersion 
(for a sub sample), the K-correction, and the most prominent spectral indicators 
are given.

Comparison of LRS results with IRS and literature data indicate that low 
resolution spectra can provide accurate values for the redshift (within  65~\kms),
and for the Mg$_2$ strength (within  0.01 mag). Moreover, since the high and 
moderate resolution measurements are tightly related, a well defined correction
can be applied to obtain all the data in a common base. This  means that accurate 
enough data can be obtained with moderate resolutions, and therefore for galaxies 
to rather high redshift, an important point for any evolutionary study.

Our data have allowed to give more precise values for the redshift and velocity
dispersion  of 4 clusters, poorly  known to now, namely A98, A3125, A3330, and
DC2103 (see Table~\ref{rescum}).


Regarding the K-correction terms, our results indicate that it is very similar 
for all the bright E/S0 galaxies in a given cluster, so the same correction can
be applied to all for many studies. Our values are in very good agreement with 
the data by Pence (1977) and with the model predictions by Poggianti (1997).

Given the short range in redshift of the clusters analyzed here, no trend with z 
was expected for any of the spectral indicators, as it is found. It is 
interesting to note, however, that for all three spectral indicators we find a
large range of the values for every cluster. It is that physical variance in the
spectral properties of the member galaxies what seems to be the most interesting 
aspect from the present analysis.
%
%
A trend between the line indicators NaD and Mg$_2$  is present in our data, in
agreement  with  previous results. The 4000\AA~Break  doesn't seem to have a 
clear relation with the other indicators.
%

 \begin{table*}
      \caption[]{Galaxies not included in Dressler (1980)}
         \label{pos}
$$
         \begin{array}{lcccrccccc}
            \hline
{\rm Galaxy} & {\rm \alpha (2000)} & {\rm \delta (2000)} & {Spectrum~id.}\\
\hline
Abell~98 & & \\
\hline
2MASXJ00463567+2029428 &  00:46:35.7 & $+$20:29:42 & F1g20\\
2MASXJ00463524+2030108  &  00:46:35.3 & $+$20:30:11& F1g22\\
A98:[BGH82]320 &  00:46:24.7 & $+$20:30:07 	   & F1g30 \\
2MASXJ00462380+2030006 &  00:46:23.9 & $+$20:30:00 & F1g31\\
2MASXJ00462059+2029076  &  00:46:20.7 & $+$20:29:07& F1g38\\
2MASXJ00461931+2029416 &  00:46:19.4 & $+$20:29:41 & F1g43\\
A98:[BGH82]311 &  00:46:16.0 & $+$20:30:19	   & F1g44 \\
2MASXJ00463702+2026068 &  00:46:37.1 & $+$20:26:06 & F1g58\\
2MASXJ00462586+2027326 &  00:46:25.9 & $+$20:27:32 & F1g65\\
3C 021    &  00:46:29.4 & $+$20:28:04   	   & F1g76\\
2MASXJ00463318+2027518 &  00:46:33.2 & $+$20:27:51 & F1g77\\
2MASXJ00463183+2028118 &  00:46:31.9 & $+$20:28:11 & F1g80\\
2MASXJ00463624+2028268 &  00:46:36.3 & $+$20:28:26 & F1g88\\
2MASXJ00463624+2028268 &  00:46:08.5 & $+$20:28:50 & F2g01\\
2MASXJ00460743+2028486 &  00:46:07.5 & $+$20:28:48 & F2g02\\
A98:[PBL2000]0264 &  00:46:04.8 & $+$20:28:27      & F2g03 \\
2MASXJ00455043+2027487 &  00:45:50.5 & $+$20:27:49 & F2g05\\
2MASXJ00455043+2027487  &  00:45:50.4 & $+$20:29:08 & F2g07A\\
2MASXJ00455008+2029097 &  00:45:50.2 & $+$20:29:09  & F2g07B\\
2MASXJ00460209+2030516 &  00:46:02.2 & $+$20:30:50 & F2g09\\
2MASXJ00461355+2034516 &  00:46:13.6 & $+$20:34:51 & F3g01\\
2MASXJ00460400+2034516 &  00:46:04.0 & $+$20:34:51 & F3g02\\
A98:[BGH82]182 &  00:45:59.8 & $+$20:35:09 	   & F3g03\\
NPM1G+20.0024 &  00:45:52.9 & $+$20:35:13 	   & F3g04\\
A98:[PBL2000]376&  00:45:57.5 & $+$20:36:56 	   & F3g05\\
2MASXJ00460400+2034516&  00:46:04.5 & $+$20:36:45  & F3g07\\
A98:[PBL2000]269 &  00:46:17.1 & $+$20:23:44 	   & F10g5\\
2MASXJ00461469+2023426 &  00:46:14.7 & $+$20:23:42 & F10g6\\
2MASXJ00462849+2023488 &  00:46:28.5 & $+$20:23:48 & F10g7\\
2MASXJ00463390+2023548 &  00:46:34.0 & $+$20:23:55 & F10g8\\
2MASX J00463852+2022538 &  00:46:38.6 & $+$20:22:53& F10g9\\
\hline
Abell~3330 & & \\
\hline
2MASXJ05145263-4859071   &  05:14:52.5 & $-$48:59:07 & g21 \\
2MASXJ05145588-4859121 &  05:14:55.7 & $-$48:59:12 & g22\\
2MASXJ05150675-4901321  &  05:15:06.6 & $-$49:01:33 & g5\\
2MASXJ05150726-4902261  &  05:15:07.2 & $-$49:02:27 & g6\\
APMUKS(BJ)B051342.65-490606.2 &  05:15:00.7 & $-$49:02:48 & g7\\
FAIRALL 0790  &  05:14:39.4 & $-$49:03:29  & g1 \\
2MASXJ05143494-4905153  &  05:14:34.9 & $-$49:05:15 & g2\\
2MASXJ05150574-4905461  &  05:15:05.6 & $-$49:05:47 & g4\\
2MASX J05142945-4905534  &  05:14:29.3 & $-$49:05:54 & g3\\

        \noalign{\smallskip}\hline\noalign{\medskip}
 \end{array}
     $$
   \end{table*}
{\scriptsize
\clearpage \onecolumn
\begin{longtable}{l|rcrrrrcccl|cccccc}
    \caption{\label{res}Observed parameters of the standard and cluster 
galaxies}\\
      \hline
{\rm Galaxy} & \multicolumn{10}{c}{\rm LRS data} & \multicolumn{6}{c}
{\rm IRS data}\\
         & {\rm cz} & {\rm Q} & {\rm Refs} & {\rm k$_B$} & {\rm k$_V$} &
{\rm k$_r$} & {\rm D4000} & {\rm Mg$_2$} & {\rm NaD} & {\rm
Run} &  {\rm cz} & {\rm $\epsilon$} & {\rm $\sigma$} & {\rm $\epsilon$} &
{\rm Mg$_2$} & {\rm Comments} \\
\hline\hline
\endfirsthead
\caption{continued}\\
      \hline
{\rm Galaxy} & \multicolumn{10}{c}{\rm LRS data} & \multicolumn{6}{c}
{\rm IRS data}\\
         & {\rm cz} & {\rm Q} & {\rm Refs} & {\rm k$_B$} & {\rm k$_V$} &
{\rm k$_r$} & {\rm D4000} & {\rm Mg$_2$} & {\rm NaD} & {\rm
Run} &  {\rm cz} & {\rm $\epsilon$} & {\rm $\sigma$} & {\rm $\epsilon$} &
{\rm Mg$_2$} & {\rm Comments} \\
\hline
\endhead
\hline
\endfoot
Std Gal &   &   &   &   &   &   &    &   &   &  &   &   &   &   &   &   \\
\hline\noalign{\smallskip}
N541     & 5427 & 1 & a,~b & .08 & .02 & $-$.03 & 1.76 & .296 & 3.4 & 5 &  --  & -- & --  & -- & --   & -- \\
N720     & 1758 & 1 & a,~b & .02 & .01 & .00    &  --  & .318 & 4.5 & 2 & 1729 & 17 & 223 & 74 & .323 & -- \\
N1395    & 1699 & 1 & a,~b & --  & .01 & .00    &  --  & .316 & 4.9 & 1 & 1667 & 38 & 275 & 28 & .344 & -- \\
N1399    & 1450 & 1 & a,~c & --  & .01 & .00    &  --  & .341 & 5.3 & 2 & 1445 & 51 & 364 & 74 & .336 & -- \\
N1403    & 4292 & 1 & a,~d & --  & .02 & .00    & 1.82 & .281 & 3.2 & 2 & 4281 & 36 & 178 & 43 & .265 & -- \\
N1406    & 1357 & 1 & e    & .02 & .01 & .00    & 2.17 & .266 & 2.8 & 2 &  --  & -- & --  & -- & --   & -- \\
N1426    & 1409 & 1 & a,~f & --  & .01 & .00    &  --  &  --  & 3.9 & 2 & 1425 & 17 & 156 & 23 & .260 & $\Delta$V = 172 \\
N1726    & 3964 & 1 & a,~g & --  & --  & --     &  --  & .288 & 4.1 & 5 &  --  & -- & --  & -- & --   & -- \\
N2340    & 5971 & 1 & b    & --  & --  & --     & 2.48 & .335 & 5.8 & 3 & 5938  & 37 & 241  & 33 & .337 & -- \\
N2974    & -- & -- & --    & --  & --  & --      & --    & --     & --  & 4 & 1980  & 51 & 253  & 35 & .289 & -- \\
N2986    & 2336 & 1 & a,~b & --  & .01 & .00    &  --  & .319 & 5.3 & 1 &  --  & -- & --  & -- & --   & -- \\
E462G15  & 5751 & 1 & a,~h & --  & .04 & .01    & 1.87 & .292 & 3.9 & 2 & 5812 & 75 & 263 & 15 & .272 & -  \\
N7507    & 1538 & 1 & d    & --  & .01 & .01    &  --  & .335 & 5.0 & 2 & 1578 & 18 & 206 & 26 & .325 & -- \\
N7562    & 3583 & 1 & b    & --  & .03 & .02    & 2.47 & .294 & 4.7 & 2 & 3606 & 58 & 242 & 17 & .288 & -- \\
N7619    & 3804 & 1 & a,~b & .05 & .03 & .02    & 1.82 & .320 & 4.9 & 2 & 3801 & 41 & 275 & 35 & .312 & -- \\
%
\hline
\hline\noalign{\smallskip}
Abell~98 &   &   &   &   &   &   &    &   &   &  &   &   &   &   &   &   \\
\hline\noalign{\smallskip}
F1g20 & 32104$^a$      & 4 & -- & .57 & -- & -- & 1.92 & .207 & 1.7 & 4 & --  & -- & --  & -- & --   & -- \\
F1g22 & 30852          & 3 & -- & .60 & -- & -- & 2.24 & .187 & 3.0 & 4 & --  & -- & --  & -- & --   & -- \\
F1g30 & 31358          & 2 & -- & .56 & -- & -- & 1.82 & .185 & 3.1 & 4 & --  & -- & --  & -- & --   & -- \\
F1g31 & 29984          & 2 & -- & .60 & -- & -- & 2.23 & .243 & 2.1 & 4 & --  & -- & --  & -- & --   & -- \\
F1g38 & 31021$^b$      & 2 & -- & .60 & -- & -- & 2.12 & .231 & 2.4 & 4 & --  & -- & --  & -- & --   & -- \\
F1g43 & 33099          & 3 & -- & .61 & -- & -- & 2.19 & .216 & 1.6 & 4 & --  & -- & --  & -- & --   & -- \\
F1g44 & 30605$^c$      & 2 & -- &  -- & -- & -- & --   &  --  & --  & 4 & --  & -- & --  & -- & --   & EL \\
F1g58 & 30654          & 1 & -- &  -- & -- & -- & --   &  --  & --  & 5 & --  & -- & --  & -- & --   & EL \\
F1g65 & 31049          & 2 & -- & .62 & -- & -- & 2.08 & .258 & 2.5 & 4 & --  & -- & --  & -- & --   & -- \\
F1g76 & 30898          & 2 & -- & .57 & -- & -- & 1.93 & .283 & 4.9 & 5 & --  & -- & --  & -- & --   & -- \\
F1g77 & 42272$^d$      & 2 & -- & .55 & -- & -- & 1.51 & .178 & 3.0 & 4 & --  & -- & --  & -- & --   & -- \\
F1g80 & 32312          & 2 & -- & .57 & -- & -- & 1.91 & .226 & 2.4 & 5 & --  & -- & --  & -- & --   & -- \\
F1g88 & 32122          & 1 & -- & .61 & -- & -- & 1.93 & .258 & 3.9 & 4,~5& --& -- & --  & -- & --   & -- \\
F2g1  & 30484          & 3 & -- & .40 & -- & -- & 1.45 & .119 & 2.5 & 5 & --  & -- & --  & -- & --   & H$\beta$~abs.\\
F2g1A & 31475          & 2 & -- & .51 & --  & -- & 1.92 & .201 & 2.6 & 5 & --  & -- & --  & -- & --   & -- \\
F2g3   & 31605          & 2 & -- & .52 & --  & -- & 1.83 & .181 & 2.9 & 5 & --  & -- & --  & -- & --   & -- \\
F2g5   & 31100          & 2 & -- & .51 & --  & -- & 1.85 & .239 & 2.3 & 5 & --  & -- & --  & -- & --   & -- \\
F2g7A & 31005         & 3 & -- & .51 & --  & -- & 1.85 & .243 & 3.4 & 5 & --  & -- & --  & -- & --   & -- \\
F2g7B & 30510         & 1 & -- & .52 & --  & -- & 1.68 & .218 & 1.8 & 5 & --  & -- & --  & -- & --   & -- \\
F2g9   & 30151         & 3 & -- & .53 & --  & -- & 1.92 & .238 & 3.6 & 5 & --  & -- & --  & -- & --   & -- \\
F3g1   & 17871         & 1 & -- & .29 & -- & -- & 1.89 & .279 & 2.9 & 5 & --  & -- & --  & -- & --   & -- \\
F3g2   & 30169         & 2 & -- & .57 & --  & -- & 1.76 & .233 & 2.6 & 5 & --  & -- & --  & -- & --   & -- \\
F3g3  & 31164$^e$  & 2 & -- & .48 & -- & -- & 1.76 & .217 & 3.3 & 5 & --  & -- & --  & -- & --   & -- \\
F3g4  & 10988      & 1 & -- & --   & -- & -- & --   &  --  & --  & 5 & --  & -- & --  & -- & --   & EL \\
F3g5  & 30343      & 2 & -- & .41 & -- & -- & 1.35 & .157 & 2.3 & 5 & --  & -- & --  & -- & --   & -- \\
F3g7  & 31284      & 2 & -- & .49 & -- & -- & 1.85 & .261 & 2.0 & 5 & --  & -- & --  & -- & --   & -- \\
F10g5 & 35868      & 1 & -- & .55 & -- & -- & 1.82 & .205 & 1.7 & 5 & --  & -- & --  & -- & --   & -- \\
F10g6 & 31037      & 1 & -- & .52 & -- & -- & 1.87 & .211 & 2.4 & 5 & --  & -- & --  & -- & --   & -- \\
F10g7 & 30370      & 1 & -- & .49 & -- & -- & 1.68 & .174 & 3.6 & 5 & --  & -- & --  & -- & --   & -- \\
F10g8 & 32834      & 1 & -- & .52 & -- & -- & 1.76 & .201 & 2.4 & 5 & --  & -- & --  & -- & --   & -- \\
F10g9 & 31094      & 1 & -- & .45 & -- & -- & 1.53 & .161 & 2.1 & 5 & --  & -- & --  & -- & --   & -- \\
\hline
\hline\noalign{\smallskip}
Abell~119 &   &   &   &   &   &   &    &   &   &  &   &   &   &   &   &   \\
\hline\noalign{\smallskip}
D26  & 13539 & 1 & --            & --  & .08 & .03   & 1.87 & .259 & 3.8 & 2 & 13462 & 24 & 173 & 66 & .246 & -- \\
D36  & 13958 & 2 & --            & .26 & .04 & .03   & 1.93 & .269 & 2.7 & 4 & --    & -- & --  & -- & --   & -- \\
D37  & 12959 & 1 &1,3,4       & --  & .08 & .03   &  --  & .319 & 4.8 & 1 & 12872 & 26 & 240 & 30 & .309 & $\Delta$ V = 55 \\
D38  & 12591 & 1 &  1            & .22 & .04 &  --   & 1.88 & .269 & 2.5 & 4 & --    & -- & --  & -- & --   & -- \\
D41  & 12408 & 1 & 1,3        & --  & .07 & .03   &  --  &  --  & 3.9 & 1 & 12342 & 38 & 173 & 22 & .316 & $\Delta$ V = 250 \\
D44$^f$  & 13184 & 1 & 1           & --  & .06 & .02   &  --  & .298 & 3.0 & 1 & 13081 & 66 & 201 & 35 & .315 & EL \\
D45  & 12660 & 1 & --          & --  & --  & --   & --   &  --  & --  & 1 & --    & -- & --  & -- & --   & EL \\
D47  & 14521 & 3 & 1          & .22 & .10 &  .04  & 1.27 & .281 & 2.8 & 2 & 14628 & 40 & 194 & 37 & .261 & $\Delta$ V = 186 \\
D49  & 13810 & 2 & 1          & --  & .09 & .05 & 1.62 & .219 & 2.5 & 2 & 13733 & 67 & 291 & 43 & .233 & -- \\
D51  & 12540 & 1 & 1          & .24 & .03 & --    & 1.96 & .273 & 4.2 & 4 & --    & -- & --  & -- & --   & -- \\
D52  & 13413 & 1 & 1,2,3,5 & --  & .08 & .03   & 1.81 &  --  & 4.4 & 1,~2& 13447 & 50 & 264 & 30 & .331 & -- \\
D60  & 11475 & 2 & 1,2,3,6& --  & .06 & .02   &  --  & .288 & 4.1 & 1 & 11565 & 56 & 331 & 25 & .302 & -- \\
D62  & 12990 & 3 & 1      & --  & .07 & .03   &  --  & .313 & 3.4 & 1 & 13077 & 28 & 161 & 19 & .321 & $\Delta$ V = 70 \\
D66  & 13318 & 1 & 1,2,3,6& --  & .07 & .02 &  --  & .295 & 4.8 & 1 & 13356 & 50 & 270 & 55 & .317 & -- \\
D68  & 12143 & 2 & 1      & --  & .07 & .02   &  --  & .264 & 4.4 & 1 & 12225 & 44 & 220 & 42 & .308 & $\Delta$ V = 380 \\
D74$^g$ & 12663 & 1 & 1      & --  & .05 & .01   &  --  & .252 & 3.5 & 1 & 12647 & 61 & 249 & 28 & .240 & -- \\
D75$^h$ & 11532 & 1 & 1,5    & --  & .05 & .01   &  --  & .261 & 2.7 & 1,~2& 11560 & 66 & 167 & 31 & .256 & -- \\
D93$^i$ & 11703 & 2 & 1,5    & --  & .05 & .02   &  --  & .297 & 4.5 & 1 & 11705 & 48 & 210 & 72 & .272 & -- \\
D94  & 12553 & 1 & 1,5    & .23 & .04 & --    & 2.03 & .264 & 3.0 & 4 & --    & -- & --  & -- & --   & -- \\
D99$^j$ & 13698 & 2 & 1      & --  & .07 & .05   &  --  & --  & 2.8 & 1,~2& 13540& 59 & 268 & 34 & .288 & -- \\
D102 & 13230 & 1 & 1      & --  & .07 & .04   & 2.00 &.161 & 4.8 & 1,~2& 13258& 61 & 196 & 48 & .178 & -- \\
D105 & 13335 & 1 & 1,2,3  & --  & .08 & .03   & 1.67 & --  & 3.7 & 1 & 13350 & 62 & 313 & 89 & .286 & -- \\
D107$^k$ & 13184 & 2 & 1       & .16 & .06 & .01   & 1.00 & --  & 3.4 & 2 & 13085& 68 & 157 & 35 & .316 & $\Delta$ V = 107 \\
D109$^l$ & 13029 & 1 & 1,5     & --  & .05 & .01   & --   &.202 & 3.3 & 1,~2& 13021 & 57& 224 &  8 & .195 &  -- \\
D111 & 12615 & 1 & --      & .20 & .08 & .04   & 1.55 &.262 & 4.2 & 2 & 12585 & 70 & 249 & 54 & .255 & -- \\
D112$^m$ & 14560 & 3 & 1       & .19 & .06 & 0.02 & 1.43 &.196 & 2.6 & 2 & 14650 & 72 & 219 & 45 & .227 & -- \\
D114$^n$ & 13312 & 2 & 1       & --  & .07 &.03 & 2.00 &.288 & 3.9 & 2 & 13375 & 40 & 128 & 11 &  --  & $\Delta$ V = 90 \\
\hline\noalign{\smallskip}
Abell~3125  &   &   &   &   &   &   &    &   &   &  &   &   &   &   &   &   \\
\hline\noalign{\smallskip}
D9   & 18508 & 1 & --  & -- & .14   & .08  & 1.45 & .261 &  -- & 1 & --    & -- & --  & -- & --   & -- \\
D11  &  8682$^o$& 1 & 1,8 & -- & .02 & .03 &  --  &  --  &  -- & 1 & --    & -- & --  & -- & --   & EL \\
D14  & 15532 & 1 & --  & -- & .15    & .12   &  --  &  --  &  -- & 1 & --    & -- & --  & -- & --   & EL \\
D46  & 17900 & 3 &  9  & -- & .13    & .08   & 1.64 & .270 & 4.4 & 1 & 17830 & 38 & 253 & 24 & .276 & $\Delta$ V = 50 \\
D47  & 17816 & 2 & 9   & -- & .13    & .09   & 1.76 & .241 & 3.8 & 1 & 17745 & 31 & 299 & 42 & .269 & -- \\
D48  & 17346 & 1 & 9   & -- & .12    & .08   & 2.14 & .189 & 3.0 & 1 & --    & -- & --  & -- & --   & -- \\
D51  & 18790 & 1 & --  & -- & .14    & .09   & 2.08 & .282 & 4.3 & 1,~2&18845& 36 & 213 & 46 & .300 & -- \\
D60  & 17587 & 1 & 9   & -- & .13    & .09   & 1.54 & .255 & 3.6 & 1 & 17478 & 45 & 291 & 59 & .281 & -- \\
D77  & 17440 & 2 & --  & -- & .11    & .07   &  --  &  --  &  -- & 1 & 17298 & 70 & 261 & 67 &  --  &  EL \\
D88  & 17675 & 2 & --  & -- & .14    & .09  & 1.06 & .332 & 4.7 & 1 & 17693 & 71 & 332 & 85 & .310 & $\Delta$ V = 189 \\
D93  & 17485 & 1 & 9   & -- & .11    & .07  & 1.22 & .259 & 2.9 & 2 & 17621 & 35 & 287 & 77 & .271 & -- \\
D95  &  --   & - & --  &    &  --    & --    & --   &  --  & --  & - & 17866 & 61 & 335 & 34 & .263 & $\Delta$ V = 73 \\
D96  & 17702 & 3 & 9   & -- & .13    & .10   & 2.33 & .198 & 2.0 & 1 & 17652 & 48 & 239 & 22 & .215 & -- \\
D103 & 17786 & 1 & --  & -- & .13    & --    & 1.54 & .177 & 3.0 & 2 & 17766 & 49 & 288 & 11 & .185 & -- \\
D104 & 18099 & 1 & --  & -- & .14    & .09   & 1.67 & .237 & 2.8 & 1 & --    & -- & --  & -- & --   & -- \\
D129 & 18890 & 1 & 9   & -- & .13    & .09   & 1.47 & .231 & 2.6 & 1 & --    & -- & --  & -- & --   & -- \\
D130 & 19112 & 2 & 9   & -- & .13    & .09   & 1.53 & .281 & 3.2 & 1 & 19049 & 27 & 236 & 25 & .271 & -- \\
D140 & 18917 & 2 & 9   & -- & .15    & .10   & 1.64 & .211 & 3.9 & 1 & 18945 & 39 & 212 &  5 & .209 & -- \\
D160 & 18361 & 1 & 9   & -- & .15    & .10   & 1.39 & .263 & 3.2 & 1 & 18459 & 46 & 262 & 61 &  --  & EL \\
D161 & 18115 & 2 & --  & -- & .14    & .10   & 1.68 & .315 & 2.8 & 1,~2& 17964& 79& 328 & 35 & .336 & -- \\
\hline
\hline\noalign{\smallskip}
Abell~3330  &   &   &   &   &   &   &    &   &   &  &   &   &   &   &   &   \\
\hline\noalign{\smallskip}
G1   & 27522 & 2 & -- & -- & .10 & --  & 1.54 & .244 & 4.0 & 1 & --    & -- & --  & -- & --   & -- \\
G2   & 26446 & 2 & -- & -- & .10 & -- & 1.00 & .226 & 1.6 & 1 & --    & -- & --  & -- & --   & -- \\
G3   & 27358 & 2 & -- & -- & .15 & -- & 2.66 & .260 & 3.2 & 1 & --    & -- & --  & -- & --   & -- \\
G4   & 26586 & 3 & -- & -- & .13 & -- & 1.97 & .242 & --  & 1 & --    & -- & --  & -- & --   & -- \\
G5   & 28262 & 1 & -- & -- & .17 & --  & 1.45 & .314 & 5.5 & 1 & --    & -- & --  & -- & --   & -- \\
G6   & 27140 & 1 & -- & -- & .14 & --  & 1.68 & .258 & 3.8 & 1 & --    & -- & --  & -- & --   & -- \\
G7   & 26222 & 1 & -- & -- & .11 & --  & --   & --   & --  & 1 & --    & -- & --  & -- & --   & EL \\
G21  & 26463 & 1 & -- & -- & .19 & --  & 1.50 & .324 & 2.9 & 1 & --    & -- & --  & -- & --   & -- \\
G22  & 28251 & 1 & -- & -- & .20 & --  & 1.44 & .247 & 3.0 & 1 & --    & -- & --  & -- & --   & -- \\
\hline
\hline\noalign{\smallskip}
Abell~1069  &   &   &   &   &   &   &    &   &   &  &   &   &   &   &   &   \\
\hline\noalign{\smallskip}
D3    & 19336 & 1 & -- & --   & .10 & .07 & 1.16 & .232 & 4.8 & 1 & --    & -- & --  & -- & --   & -- \\
D9    & 19478 & 3 & -- & --   & .11 & .07 & 2.44 & .234 & 3.6 & 1 & --    & -- & --  & -- & --   & -- \\
D12   & 16624 & 2 & 7  & --   & .08 & .07 & 1.66 & .253 & 2.3 & 1 & --    & -- & --  & -- & --   & -- \\
D14   & 16656 & 1 & -- & --   & .08 & .07 & 1.38 & .275 & 3.6 & 1 & --    & -- & --  & -- & --   & -- \\
D15   & 19737 & 1 & 1,7& --   & .10 & .07 & 1.51 & .235 & 3.9 & 1,6 & 19709 & 23 & 239  & 9 & .292 & -- \\
D16   & 20110 & 4 & 1  & --   & .09 & .07 & 1.00 & .140 & 5.2 & 1 & --    & -- & --  & -- & --   & -- \\
D19   & 18030 & 1 & 1,7& --   & .10 & .08 & 1.10 & .224 & 3.7 & 1,6 & 18040 & 31 & 220  & 11 & .261 & -- \\
D21   & 19817 & 2 & 1,7& --   & .11 & .06 & 1.20 & .319 & 3.2 & 1,6 & 19345 & 31 & 357  & 9 & .328  & -- \\
D25   & 20331 & 1 & 1,7& --   & .09 & .04 & 1.82 & .212 & 2.3 & 1,6 & 20287 & 19 & 275  & 7 & .219   & -- \\
D29   & 20051 & 1 & 1,7& --   & .10 & .07 & 2.00 & .259 & 4.5 & 1,6 & 20067 & 17 & 236  & 7 & .283   & -- \\
D30   & 20430 & 1 & 1  & --   & .08 & .06 & 1.22 & .219 & 3.4 & 1 & --    & -- & --  & -- & --   & -- \\
D38   & 19329 & 2 & -- & --   & .11 & .08 &  --  & .244 &  -- & 1,6 & 19432 & 19 & 295  & 7 & -- & -- \\
D39   & 19729 & 3 & -- & --   & .09 & .07 &  --  & .192 & 1.7 & 1 & --    & -- & --  & -- & --   & -- \\
D41   & 11644 & 1 & -- & --   & --  & --  &  --  &  --  & --  & 1 & --    & -- & --  & -- & --   & EL \\
D46   & 19700 & 2 & -- & --   & .09 & .05 & 1.97 & .248 & 4.8 & 1,6 & 19653 & 19 & 231  & 8 & .307   & -- \\
D47   & -- & - & -- & --   & -- & -- & -- & -- & -- & 6 & 18931 & 14 & 149  & 10 & .250   & -- \\
\hline
\hline\noalign{\smallskip}
Abell~1983  &   &   &   &   &   &   &    &   &   &  &   &   &   &   &   &   \\
\hline\noalign{\smallskip}
D6   & 17547 & 1 & 10 & -- & --  & -- & 1.66 & .186 & 2.0 & 3 & --    & -- & --  & -- & --   & -- \\
D10  & 13156 & 1 & 10 & -- & .07 & -- & 1.60 & .303 & 5.3 & 3 & --    & -- & --  & -- & --   & -- \\
D15  & 13589 & 1 & 10 & -- & .07 & -- & 2.11 & .292 & 4.3 & 3,6 & 13676 & 20 & 323  & 7 & .323   & -- \\
D23  & 12782 & 1 & 10 & -- & .05 & -- & 1.46 & .154 & 2.3 & 3 & --    & -- & --  & -- & --   & -- \\
D24  & 13504 & 1 & 10 & -- & .07 & -- & 1.59 & .241 & 4.4 & 3 & --    & -- & --  & -- & --   & -- \\
D26  & 17822 & 1 & 10 & -- & --  & -- & 1.90 & .294 & 4.4 & 3 & --    & -- & --  & -- & --   & -- \\
D28  &  6251 & 1 & 10 & -- & .04  & -- & 1.59 & .180 & 2.0 & 3 & --    & -- & --  & -- & --   & -- \\
D35  & 13659 & 1 & 10 & -- & .08 & -- & 2.36 & .237 & 4.1 & 3 & --    & -- & --  & -- & --   & -- \\
D46  & 13532 & 1 & 10 & -- & .07 & -- & 2.15 & .280 & 2.7 & 3 & --    & -- & --  & -- & --   & -- \\
D54  & 13167 & 2 &  7 & -- & .07  & -- & 1.79 & .297 & 4.7 & 3,6 & 12979 & 14 & 232  & 7 & .298 & -- \\
D56  & 13481 & 1 & 10 & -- & .07 & -- & 2.04 & .285 & 4.8 & 3,6 & 13583  & 20 & 198  & 9 & .265 & -- \\
D58  & 13606 & 1 & 11 & -- & .07 & -- & 2.03 & .212 & 2.6 & 3 & --    & -- & --  & -- & --   & -- \\
D77  & -- & - &  -- & -- & -- & -- & -- & -- & -- & 6 & 13844 & 15 & 215 & 8 & .266 & -- \\
D78  & 13781 & 1 &  7 & -- & .06 & -- & 1.85 & .306 & 3.9 & 3,6 & 13639 & 15 & 212 & 8 & .277 & -- \\
D84  & 13894 & 1 & 11 & -- & .08 & -- & 1.96 & .281 & 4.8 & 3 & --    & -- & --  & -- & --   & -- \\
D105 & 13310 & 1 & 10 & -- & .06 & -- & 1.98 & .264 & 4.3 & 3,6 & 13387 & 20 & 323 & 7 & .323 & -- \\
\hline
\hline\noalign{\smallskip}
Abell~2151  &   &   &   &   &   &   &    &   &   &  &   &   &   &   &   &   \\
\hline\noalign{\smallskip}
D4   &  9986 & 3 & 10 & .18 & .06 & -- & 1.91 & .219 & 1.9 & 4 & --    & -- & --  & -- & --   & -- \\
D7   & 10055 & 1 &  2 & .20 & .07 & -- & 2.23 & .290 & 3.2 & 4 & --    & -- & --  & -- & --   & -- \\
D9   & 10248 & 1 & 10 & .18 & .06 & -- & 1.90 & .261 & 1.7 & 4 & --    & -- & --  & -- & --   & -- \\
D15  &  9966 & 1 & 10 & .19 & .06& -- & 1.96 & .281 & 2.6 & 4 & --    & -- & --  & -- & --   & -- \\
D40  & 10330 & 1 & 12 & .19 & .04& -- & 2.12 & .268 & 3.1 & 4,6 & 10050  & 15 & 230  & 7 & .271 & -- \\
D62  &  9338 & 1 &  2 & .18 & .06 & -- & 2.36 & .233 & 3.2 & 4 & --    & -- & --  & -- & --   & -- \\
D63  & 10830 & 1 & 10 & .21 & .07 & -- & 2.43 & .281 & 3.1 & 4,6 & 10455 & 16 & 195 & 9 & .262 & -- \\
D64  & 10432 & 2 & 10 & .21 & .07 & -- & 1.97 & -- & 5.0 & 4 & --    & -- & --  & -- & --   & -- \\
D65  & -- & - & -- & -- & -- & -- & -- & -- & -- & 6 & 10426  & 13 & 291 & 6 & .305   & -- \\
D66  & -- & - & -- & -- & -- & -- & -- & -- & -- & 6 & 9643  & 9 & 186 & 6 & .270   & -- \\
D78  & 10262 & 1 & -- & .13 & .05 & --  & --   &  --  & --  & 4 & --    & -- & --  & -- & --   & EL \\
D134 & -- & - & -- & -- & -- & -- & -- & -- & -- & 6 & 11074  & 50 & 318 & 11 & .302   & -- \\
\hline
&   &   &   &   &     &   &    &   &   &  &   &   &   &   &   &   \\
&   &   &   &   &     &   &    &   &   &  &   &   &   &   &   &   \\
\hline\noalign{\smallskip}
DC~2103-39  &   &   &   &   &     &   &    &   &   &  &   &   &   &   &   &   \\
\hline\noalign{\smallskip}
D3   & 16114 & 1 & -- & .26 & .13 & .03 & 1.33 & .268 & 4.2 & 2 & 16152 & 35  & 185  &  27 & .272 & $\Delta$ V = 150 \\
D14  &  9374 & 1 & -- & .13 & .05 & .02 & 1.28 & .329 & 4.9 & 2 &  9343 & 40  & 332  &  58 & .327 & -- \\
D15  & 15034 & 1 & -- & .23 & .09 & .04 & 1.32 & .335 & 3.2 & 2 & 15038 & 39  & 237  &  30 & .316 & $\Delta$ V = 103 \\
D18  &  9319 & 2 & -- & --  & .05 & .03 &  --  &  --  & --  & 2 &  9208 & 30  & 137  &  38 &  --  & EL,~$\Delta$ V = 122 \\
D20  &  9287 & 1 & -- & .11 & .04 & .02 & 1.00 & .185 & 1.8 & 2 &  9353 & 19  & 154  &  38 & .191 & -- \\
D21  &  9161 & 1 & -- & .15 & .06 & .03 & 1.49 & .229 & 2.5 & 2 &  9400 & 31  & 242  &  64 & .218 & $\Delta$ V = 91 \\
D23  &  9279 & 1 & -- & .13 & .07 & .03 & 1.25 & .273 & 2.6 & 2 &  9245 & 62  & 260  &  52 & .244 & $\Delta$ V = 270 \\
D38  & 15820 & 1 & -- & .22 & .08 & .04 & 1.60 & .232 & 3.0 & 2 & 15742 & 52  & 275  &  47 & .212 & -- \\
D39  & 15869 & 3 & -- & .26 & .09 & .04 & 1.19 & .252 & 1.0 & 2 & 15892 & 51  & 313  &  23 & .268 & -- \\
D40  & 15060 & 1 & -- & .22 & .08 & .04 & 1.47 & .249 & 3.9 & 2 & 14992 & 33  & 223  &  17 & .260 & -- \\
D42  & 14719 & 1 & -- & .20 & .10 & --  & 1.47 & .240 & 3.6 & 2 & 14700 & 24  & 206  &  71 & .249 & -- \\
D53  & 26506 & 1 & -- & --  & .13 & --  & 1.97 & .287 & 3.7 & 2 & 26558 & 60  & 316  &  23 &  --  & -- \\
D60  & 15635 & 2 & -- & .22 & .11 & .05 & 1.25 & .224 & 3.7 & 2 & 15626 & 32  & 177  &  58 & .232 & -- \\
D61  & 15369 & 2 & -- & .20 & .10 & .04  & 1.33 & .278 & 3.1 & 2 & 15326 & 33  & 200  &  69 &  --  & -- \\
D62  & 15149 & 1 & -- & --  & .07 & .03 & 1.07 & .227 & 1.4 & 2 & 15079 & 23  & 176  &  17 & .230 & -- \\
D63  & 15003 & 2 & -- & --  & .08 & .03 & 1.16 &  --  & 1.9 & 2 & 14937 & 23  & 147  &  24 & .271 & -- \\
D66  &  9527 & 2 & -- & --  & .04 & .02 & 1.18 & .166 & 0.8 & 2 &  9581 & 39  & 235  &  23 & .177 & -- \\
D71  & 14836 & 1 & -- & --  & .08 & .04 & 1.37 & .249 & 4.1 & 2 & 14832 & 44  & 207  &  33 & .258 & -- \\
D73  & --    & - & -- & --  & --  & --  &  --  &  --  &  -- & - & 14940 & 38  & 260  &  17 & .276 & -- \\  
D76  & 15230 & 3 & -- & .20 & .08 & .03 & 1.66 & .187 & 4.0 & 2 & --    & --  & --   & --  & --   & -- \\
D102 & --    & - & -- & --  & --  & --  &  --  &  --  &  -- & - & 15324 & 39  & 207  &  55 &  --  & $\Delta$ V = 92 \\
\end{longtable}}
\begin{scriptsize}
References for the redshift. (A) Standard Galaxies: a, JFK; b, Smith et al (2000); c, Graham et al (1998); d, de Vaucouleurs et al (1991, RC3); 
e, Bureau et al (1996); f, Lauberts \& Valentijn (1989, ESO-Uppsala Catalog); g, Simien and Prugniel (1997); h: da Costa et al. (1991). (B) Cluster Galaxies. 1: Katgert et al (1998, ENACS Catalog); 2, de Vaucouleurs et al (1991, RC3); 3, Huchra et al (1999, CfA Catalog); 4, Zabludoff et al (1993); 
5, Paturel et al (2003, LEDA Catalogue); 6, Postman \& Lauer (1995); 7, Beers et al (1991); 8, Lauberts \& Valentijn (1989, ESO-Uppsala Catalog); 9, Caldwell \& Rose (1997); 10, Dressler \& Shectman (1988); 11, Zabludoff et al (1990); 12,	Davoust \& Considere (1995)
{\sl Notes on discrepant redshift results.}
\begin{itemize}
\item{a. In agreement with B82, but 394~\kms higher than in Zabludoff et al (1990)}
\item{b. 634~\kms higher than in Z90}
\item{c. Our value, determined form the emission lines, agrees with Z90, but it is 735~\kms lower than in Beers et al (1982)}
\item{d. Our value agrees with Z90, and is 9826~\kms higher than in Beers et al (1982)}
\item{e. The redshift we find for this galaxy is 368~\kms higher than in Beers et al (1982), but grossly discrepant with Zabludoff et al (1990)}
\item{f. Observed in LRS and IRS modes with concordant results. Our LRS value is 337~\kms higher than in the ENACS Catalogue.}
\item{g. Observed in LRS and IRS modes with concordant results. Our LRS value is 1038~\kms lower than in the ENACS Catalogue, and very close to that reported by Fouqu\'e et al (1992)}  
\item{h. Observed twice in LRS mode and in IRS mode, with concordant results. Our LRS value is 1330~\kms lower than in the ENACS Catalogue.}
\item{i. Observed in LRS and IRS modes with concordant results. Our LRS value is 1357~\kms lower than in the ENACS Catalogue. The redshift reported by Falco et al (1999) is very close to our value, whereas that reported by Fouqu\'e et al (1992) does not agree with any of previous determinations.}
\item{j. Observed in LRS and IRS modes with concordant results. Our LRS value is 1226~\kms higher than in the ENACS Catalogue.}
\item{k. Observed in LRS and IRS modes with concordant results. Our LRS value is 1210~\kms lower than in the ENACS Catalogue. The redshift reported by Fouqu\'e et al (1992) is very close to our value.}
\item{l. Observed in LRS and IRS modes with concordant results. Our LRS value is 2086~\kms higher than in the ENACS Catalogue.}
\item{m. Observed in LRS and IRS modes with concordant results. Our LRS value is 2819~\kms higher than in the ENACS Catalogue.}
\item{n. Observed in LRS and IRS modes with concordant results. Our LRS value is 5393~\kms lower than in the ENACS Catalogue.}
\item{o. Our redshift is 9478~\kms lower than in the ENACS Catalogue, that is in agreement with the redshift given in the ESO-Uppsala catalog. Our value is from the detected emission lines.}
\end{itemize}
\end{scriptsize}

%
\begin{table*}[h]
      \caption[]{Average values for the clusters}
         \label{rescum}
         \begin{tabular}{lcrcccccc}
            \hline\noalign{\smallskip}

{\rm Cluster} & {\rm cz} & {\rm $\sigma$} & {\rm Mg$_2$} & {\rm NaD} & {\rm D4000}
& {\rm k$_B$} & {\rm k$_V$} & {\rm k$_r$} \\
\hline\noalign{\smallskip}
A98   & 31225 & 895 & 0.182-0.346~(0.280) & 3.1-6.4~(4.0) & 1.45-2.34~(1.95) & 0.52 (0.06) & --  & -- \\
A119  & 13260 & 773 & 0.195-0.348~(0.303) & 2.7-5.3~(4.1) & 1.03-2.06~(1.84) & 0.22 (0.03) & 0.07 (0.02) & 0.03 (0.01) \\
A3125  & 17898 & 779 & 0.208-0.358~(0.294) & 2.3-5.0~(3.5) & 1.12-2.39~(1.64) &  --   & 0.13 (0.01)  & 0.09 (0.01) \\
A3330  & 27000 & 695 & 0.267-0.365~(0.299) & 2.1-6.0~(4.0) & 1.09-2.75~(1.63) &  -- & 0.14 (0.03) & -- \\
A1069 $^a$ & 19671 &  659 & 0.175-0.354~(0.269) & 2.0-5.5~(3.9) & 1.06-2.50~(1.44) & -- & 0.10 (0.01) & 0.07 (0.01) \\
A1983  & 13080 & 948 & 0.182-0.334~(0.309) & 2.5-5.5~(4.5) & 1.49-2.39~(1.99) & -- & 0.07 (0.01) & -- \\
A2151  & 10980 & 716 & 0.264-0.415~(0.320) & 2.7-6.0~(4.1) & 1.91-2.44~(2.06) & 0.19 (0.01) & 0.06 (0.01) & -- \\
DC2103& 15268 & 449 & 0.220-0.338~(0.282) & 1.3-4.5~(3.7) & 1.11-1.70~(1.37) & 0.22 (0.02) & 0.09 (0.02) & 0.04 (0.01) \\
\noalign{\smallskip}\hline\noalign{\medskip}
\end{tabular}
\begin{scriptsize}
The values of cz and $\sigma$for A98, A3125, A3330, and DC2103 are from the 
present work. For the
other clusters they have been taken from Struble and Rood (1999). 

For the spectral line indicators we give the range and the median value 
(in parenthesis). For the K-correction only the median
value is given.
\end{scriptsize}

      \end{table*}

\clearpage \onecolumn
\begin{longtable}{lrcrrr}
 \caption{\label{finalres}The adopted, fully corrected, redshift and spectral 
indicators}\\
 \hline\hline
{\rm Galaxy} & {\rm cz} & {\rm D4000} & {\rm NaD} & {\rm Mg$_2$} & log$\sigma$\\
\hline
\endfirsthead
\caption{continued}\\
 \hline\hline
{\rm Galaxy} & {\rm cz} & {\rm D4000} & {\rm NaD} & {\rm Mg$_2$} & log$\sigma$\\
\hline
\endhead
\hline
\endfoot
Abell~98 &  &  &  &  & \\
\hline\noalign{\smallskip}
F1g20 & 32104 & 2.02 & 3.2 &.270 & -- \\
F1g22 & 30852 & 2.34 & 4.5 &.250 & --  \\
F1g30 & 31358 & 1.92 & 4.6 &.248 & --  \\
F1g31 & 29984 & 2.33 & 3.6 &.306 & -- \\
F1g38 & 31021 & 2.22 & 3.9 &.294 & -- \\
F1g43 & 33099 & 2.29 & 3.1 &.279 & -- \\
F1g44 & 30605 & --   & --  & --  & -- \\
F1g58 & 30654 & --   & --  & --  & -- \\
F1g65 & 31049 & 2.18 & 4.0 &.321 & -- \\
F1g76 & 30898 & 2.03 & 6.4 &.346 & -- \\
F1g77 & 42272 & 1.66 & 4.6 &.242 & -- \\
F1g80 & 32312 & 2.01 & 3.9 &.289 & -- \\
F1g88 & 32122 & 2.03 & 5.4 &.321 &--   \\
F2g1  & 30484 & 1.55 & 4.0 &.182 &--  \\
F2g1A & 31475 & 2.02 & 4.1 &.264 & -- \\
F2g3  & 31605 & 1.93 & 4.4 &.244 & -- \\
F2g5  & 31100 & 1.95 & 3.8 &.302 & -- \\
F2g7A & 31005 & 1.95 & 4.9 &.306 & -- \\
F2g7B & 30510 & 1.78 & 3.3 &.281 & -- \\
F2g9  & 30151 & 2.02 & 5.1 &.301 & -- \\
F3g1  & 17871 & 1.93 & 4.1 &.332 & -- \\
F3g2  & 30169 & 1.86 & 4.1 &.296 & -- \\
F3g3  & 31164 & 1.86 & 4.8 &.280 & -- \\
F3g4  & 10988 & --   & --  & --  & -- \\
F3g5  & 30343 & 1.45 & 3.8 &.220 & -- \\
F3g7  & 31284 & 1.95 & 3.5 &.324 &  -- \\
F10g5 & 35868 & 1.92 & 3.2 &.268 & --  \\
F10g6 & 31037 & 1.97 & 3.9 &.274 & --  \\
F10g7 & 30370 & 1.78 & 5.1 &.237 & --  \\
F10g8 & 32834 & 1.86 & 3.9 &.264 & --  \\
F10g9 & 31094 & 1.63 & 3.6 &.224 & --  \\
\hline\noalign{\smallskip}
Abell~119 &  &  &  &  & \\
\hline\noalign{\smallskip}
D26  & 13462 & 1.90 & 4.0 &.263 & 2.238 \\
D36  & 13958 & 1.96 & 3.8 &.317 & --  \\
D37  & 12872 &  --  & 5.0 &.326 & 2.380 \\
D38  & 12591 & 1.91 & 3.6 &.317 & --  \\
D41  & 12342 &  --  & 4.1 &.333 & 2.238  \\
D44  & 13081 &  --  & 3.2 &.332 & 2.303 \\
D45  & 12660 &  --  &  -- & --  &  -- \\
D47  & 14628 & 1.30 & 3.0 &.278 & 2.288 \\
D49  & 13733 & 1.65 & 2.7 &.250 & 2.464 \\
D51  & 12540 & 1.99 & 5.3 &.321 & --  \\
D52  & 13447 & 1.84 & 4.6 &.348 & 2.422 \\
D60  & 11565 &  --  & 4.3 &.319 & 2.520 \\
D62  & 13077 &  --  & 3.6 &.338 & 2.207 \\
D66  & 13356 &  --  & 5.0 &.334 & 2.431 \\
D68  & 12225 &  --  & 4.6 &.325 & 2.342 \\
D74  & 12647 &  --  & 3.7 &.257 & 2.396 \\
D75  & 11560 &  --  & 2.9 &.273 & 2.223 \\
D93  & 11705 &  --  & 4.7 &.289 & 2.322 \\
D94  & 12553 & 2.06 & 4.1 &.312 &--  \\
D99  & 13540 &  --  & 3.0 &.305 & 2.428 \\
D102 & 13258 & 2.03 & 5.0 &.195 & 2.292 \\
D105 & 13350 & 1.70 & 3.9 &.303 & 2.496 \\
D107 & 13085 & 1.03 & 3.6 &.333 & 2.196 \\
D109 & 13021 &  --  & 3.5 &.212 & 2.350 \\
D111 & 12585 & 1.58 & 4.4 &.272 & 2.396 \\
D112 & 14650 & 1.46 & 2.8 &.244 & 2.340 \\
D114 & 13375 & 2.03 & 4.1 &.336 & 2.107 \\
\hline\noalign{\smallskip}
Abel~3125 &  &  &  &  & \\
\hline\noalign{\smallskip}
D9   & 18508 & 1.50 & --  &.295 &  --  \\
D11  &  8682 &  --  & --  & --  &  -- \\
D14  & 15532 &  --  & --  & --  &  -- \\
D46  & 17830 & 1.69 & 4.7 &.299 & 2.403 \\
D47  & 17745 & 1.81 & 4.1 &.292 & 2.476 \\
D48  & 17346 & 2.19 & 3.3 &.223 &  -- \\
D51  & 18845 & 2.13 & 4.6 &.323 & 2.328 \\
D60  & 17478 & 1.59 & 3.9 &.304 & 2.464 \\
D77  & 17298 & --   & --  & --  & 2.417 \\
D88  & 17693 & 1.11 & 5.0 &.333 & 2.521 \\
D93  & 17621 & 1.27 & 3.2 &.294 & 2.458 \\
D95  & 17866 & --   & --  &.287 & 2.525 \\
D96  & 17652 & 2.38 & 2.3 &.238 & 2.378 \\
D103 & 17766 & 1.59 & 3.3 &.208 & 2.459 \\
D104 & 18099 & 1.72 & 3.1 &.271 &  -- \\
D129 & 18890 & 1.52 & 2.9 &.265 &  -- \\
D130 & 19049 & 1.58 & 3.5 &.294 & 2.373 \\
D140 & 18945 & 1.69 & 4.2 &.232 & 2.326 \\
D160 & 18459 & 1.44 & 3.5 &.297 & 2.418 \\
D161 & 17964 & 1.73 & 3.1 &.358 & 2.516 \\
\hline\noalign{\smallskip}
Abel~3330 &  &  &  & & \\
\hline\noalign{\smallskip}
g1   & 27522 & 1.63 & 4.5 &.285 &  -- \\
g2   & 26446 & 1.09 & 2.1 &.267 &  -- \\
g3   & 27358 & 2.75 & 3.7 &.301 &  -- \\
g4   & 26586 & 2.06 & --  &.283 &  -- \\
g5   & 28262 & 1.54 & 6.0 &.355 &  --  \\
g6   & 27140 & 1.77 & 4.3 &.299 & --  \\
g7   & 26222 &  --  & --  & --  & --  \\
g21  & 26463 & 1.59 & 3.4 &.365 &  -- \\
g22  & 28251 & 1.53 & 3.5 &.288 &  -- \\
\hline\noalign{\smallskip}
Abell~1069 &  &  &  &  &  \\
\hline\noalign{\smallskip}
D3    & 19336 & 1.22 & 5.1 &.267 &  -- \\
D9    & 19478 & 2.50 & 3.9 &.269 & --  \\
D12   & 16624 & 1.72 & 2.6 &.288 &  -- \\
D14   & 16656 & 1.44 & 3.9 &.310 & --  \\
D15   & 19709 & 1.57 & 4.2 &.317 & 2.378 \\
D16   & 20110 & 1.06 & 5.5 &.175 &  -- \\
D19   & 18040 & 1.16 & 4.0 &.285 &  2.342 \\
D21   & 19345 & 1.26 & 3.5 &.352 & 2.553 \\
D25   & 20247 & 1.88 & 2.6 &.243 &  2,439 \\
D29   & 20067 & 2.06 & 4.8 &.307 & 2.373 \\
D30   & 20430 & 1.28 & 3.7 &.254 &  -- \\
D38   & 19432 &  --  & --  &.279 &  2.470\\
D39   & 19729 &  --  & 2.0 &.227 &  -- \\
D41   & 11644 &  --  & --  & --  &  -- \\
D46   & 19653 & 2.03 & 5.1 &.331 &  2.364 \\
D47  & 18931 &   --   &  --  & .274  & 2.173   \\
\hline\noalign{\smallskip}
Abell~1983 &  &  &  & & \\
\hline\noalign{\smallskip}
D6   & 17547 & 1.69 & 2.2 &.214 &  -- \\
D10  & 13156 & 1.63 & 5.5 &.331 & --  \\
D15  & 13676 & 2.14 & 4.5 &.340 & 2.509 \\
D23  & 12782 & 1.49 & 2.5 &.182 & --  \\
D24  & 13504 & 1.62 & 4.6 &.269 &  -- \\
D26  & 17822 & 1.93 & 4.6 &.322 & --  \\
D28  &  6251 & 1.60 & 2.1 &.205 &  -- \\
D35  & 13659 & 2.39 & 4.3 &.265 &  -- \\
D46  & 13532 & 2.18 & 2.9 &.308 & --  \\
D54  & 12979 & 1.82 & 4.9 &.315 &  2.365 \\
D56  & 13583 & 2.07 & 5.0 &.282 & 2.297 \\
D58  & 13606 & 2.06 & 2.8 &.240 &  -- \\
D77  & 13844 & -- & -- &.283 & 2.332  \\
D78  & 13639 & 1.88 & 4.1 &.294 & 2.326  \\
D84  & 13894 & 1.99 & 5.0 &.309 &  -- \\
D105 & 13387 & 2.01 & 4.5 &.340 & 2.509 \\
\hline\noalign{\smallskip}
Abel~2151 &  &  &  & & \\
\hline\noalign{\smallskip}
D4   &  9986 & 1.92 & 2.9  &.264 &  -- \\
D7   & 10055 & 2.24 & 4.2  &.335 & --  \\
D9   & 10248 & 1.91 & 2.7  &.306 & --  \\
D15  &  9966 & 1.97 & 3.6  &.326 & --  \\
D40  & 10050 & 2.13 & 4.1  &.285 & 2.362 \\
D62  &  9338 & 2.37 & 4.2  &.278 & --  \\
D63  & 10455 & 2.44 & 4.1  &.276 & 2.290 \\
D64  & 10432 & 1.98 & 6.0  & --  & -- \\
D65  & 10426 & --    & --    &.319 & 2.464 \\
D66  &  9643  & --    & --    &.284 & 2.270 \\
D78  & 10262 & --   & --   & --  &  -- \\
D134  & 11074 & --    & --    & 0.316 & 2.502 \\
\hline\noalign{\smallskip}
DC2103-39 &  &  &  & & \\
\hline\noalign{\smallskip}
D3   & 16152 & 1.37 & 4.5 &.294 & 2.267 \\
D14  &  9343 & 1.31 & 5.0 &.342 & 2.521 \\ 
D15  & 15038 & 1.36 & 3.5 &.338 & 2.375 \\
D18  &  9208 &  --  & --  & --  & 2.137 \\
D20  &  9353 & 1.01 & 1.9 &.206 & 2.188 \\
D21  &  9400 & 1.50 & 2.6 &.233 & 2.384 \\
D23  &  9245 & 1.26 & 2.7 &.259 & 2.415 \\
D38  & 15742 & 1.64 & 3.3 &.234 & 2.439 \\
D39  & 15892 & 1.23 & 1.3 &.290 & 2.496 \\
D40  & 14992 & 1.51 & 4.2 &.282 & 2.348 \\
D42  & 14700 & 1.51 & 3.9 &.271 & 2.314 \\
D53  & 26558 & 2.06 & 4.3 &.348 & 2.500 \\
D60  & 15626 & 1.29 & 4.0 &.254 & 2.248 \\
D61  & 15326 & 1.37 & 3.4 &.311 & 2.301 \\
D62  & 15079 & 1.11 & 1.7 &.252 & 2.246 \\
D63  & 14937 & 1.20 & 2.2 &.293 & 2.167 \\
D66  &  9581 & 1.19 & --  &.192 & 2.371 \\
D71  & 14832 & 1.41 & 4.4 &.280 & 2.316 \\
D73  & 14940 &  --  & --  &.298 & 2.415 \\
D76  & 15230 & 1.70 & 4.3 &.220 & -- \\
D102 & 15324 & --   & --  & --  & 2.316 \\
   \end{longtable}
  
\end{document}